\documentclass[aps,prb,12pt,a4paper,preprint,amsmath,amssymb,groupedaddress]{revtex4-1} 
\usepackage{bm}
\usepackage{graphicx}

\usepackage{mathptmx} % does not give correct varepsilon and varrho among other things
\DeclareSymbolFont{newfont}{OML}{cmm}{m}{it}% Computer Modern math font
\DeclareMathSymbol{\Varrho}{3}{newfont}{37}% Symbol 37

\newcommand{\wt}{\widetilde}

\newcommand{\ave}[1]{{\left<#1\right>}}

\newcommand{\abs}[1]{{\left|#1\right|}}

\newcommand{\rmd}{\text{d}}

\newcommand{\taud}{\ensuremath{\tau_\text{d}}}

\newcommand{\tauw}{\ensuremath{\tau_\text{w}}}

\newcommand{\Phiave}{\ensuremath{\ave{\Phi}}}
\newcommand{\Phirms}{\ensuremath{\Phi}_\text{rms}}

\newcommand{\Phiwt}{\ensuremath{\widetilde{\Phi}}}
\newcommand{\Psiave}{\ensuremath{\ave{\Psi}}}
\newcommand{\Psirms}{\ensuremath{\Psi}_\text{rms}}

\newcommand{\Psiwt}{\ensuremath{\widetilde{\Psi}}}
\newcommand{\Nwt}{\ensuremath{\widetilde{N}}}

\newcommand{\Arms}{\ensuremath{A_\text{rms}}}
\newcommand{\Aave}{\ensuremath{\ave{A}}}
\newcommand{\aveA}{\ensuremath{\ave{A}}}

\newcommand{\Refs}[1]{Refs.~\onlinecite{#1}}
\newcommand{\Eqref}[1]{Eq.~\eqref{#1}}
\newcommand{\Eqsref}[1]{Eqs.~\eqref{#1}}
\newcommand{\Figref}[1]{Fig.~\ref{#1}}
\newcommand{\Figsref}[1]{Figs.~\ref{#1}}
\newcommand{\Secref}[1]{Sec.~\ref{#1}}

\newcommand{\CPP}{\textit{Contrib.\ Plasma Phys.}}

\newcommand{\JNM}{\textit{J.~Nuclear Mater.}}

\newcommand{\JPP}{\textit{J.~Plasma Phys.}}
\newcommand{\NF}{\textit{Nucl.\ Fusion}}

\newcommand{\PFB}{\textit{Phys.\ Fluids~B}}
\newcommand{\PPCF}{\textit{Plasma Phys.\ Contr.\ Fusion}}
\newcommand{\PFR}{\textit{Plasma Fusion Res.}}

\newcommand{\PLA}{\textit{Phys.\ Lett.~A}}
\newcommand{\PP}{\textit{Phys.\ Plasmas}}

\newcommand{\PS}{\textit{Phys.\ Scripta}}

\newcommand{\BTSJ}{\textit{Bell Sys.\ Tech. J.}}

\newcommand{\PRL}{\textit{Phys.~Rev.\ Lett.}}
\newcommand{\NME}{\textit{Nucl.~Mater.\ Energy}}

\begin{document}

\title{Auto-correlation function and frequency spectrum due to a super-position of uncorrelated exponential pulses}

\author{O.~E.~Garcia}
\email{odd.erik.garcia@uit.no}
\author{A.~Theodorsen}

\affiliation{Department of Physics and Technology, UiT The Arctic University of Norway, N-9037 Troms{\o}, Norway}

\date{\today}

\begin{abstract}
The auto-correlation function and the frequency power spectral density due to a super-position of uncorrelated exponential pulses are considered. These are shown to be independent of the degree of pulse overlap and thereby the intermittency of the stochastic process. For constant pulse duration and a one-sided exponential pulse shape, the power spectral density has a Lorentzian shape which is flat for low frequencies and a power law  at high frequencies. The algebraic tail is demonstrated to result from the discontinuity in the pulse function. For a strongly asymmetric two-sided exponential pulse shape, the frequency spectrum is a broken power law with two scaling regions. In the case of a symmetric pulse shape, the power spectral density is the square of a Lorentzian function. The steep algebraic tail at high frequencies in these cases is demonstrated to follow from the discontinuity in the derivative of the pulse function. A random distribution of pulse durations is shown to result in apparently longer correlation times but has no influence on the asymptotic power law tail of the frequency spectrum. The effect of additional random noise is also discussed, leading to a flat spectrum for high frequencies. The probability density function for the fluctuations is shown to be independent of the distribution of pulse durations. The predictions of this model describe the variety of auto-correlation functions and power spectral densities reported from experimental measurements in the scrape-off layer of magnetically confined plasmas.
\end{abstract}

\maketitle

\section{Introduction}

The boundary region of magnetically confined plasmas is generally found to be in an inherently fluctuating state with order unity relative fluctuation levels of the plasma parameters.\cite{wootton,nedospasov,endler-jnm,dippolito,krasheninnikov,carreras-jnm,naulin-jnm,zweben,garcia-pfr,dmz} The frequency power spectra of these fluctuations are generally found to be characterized by a flat low-frequency region and a steep high-frequency tail.\cite{pedrosa-prl,carreras-php,rhodes,graves,antar-php1,antar-php2,dewhurst,xu-php,xu-ppcf,sattin,tanaka-nf,terry-jnm1,zweben-php-acm,terry-php,zweben-nf,zweben-php,cziegler-php} In many previous works, power law scaling relationships have been inferred from experimental measurements and the results interpreted in the frameworks of scale-free inertial range cascade, self-similar processes and self-organized critical behavior.\cite{pedrosa-prl,carreras-php,rhodes,graves,antar-php1,dewhurst,antar-php2,xu-php,sattin,tanaka-nf,xu-ppcf} By contrast, it will in this contribution be demonstrated that the auto-correlation functions and frequency spectra reported from experimental measurements can be described by a super-position of uncorrelated exponential pulses.

Large-amplitude fluctuations in the boundary region are attributed to the radial motion of blob-like plasma filaments through the scrape-off layer, most clearly demonstrated by gas puff imaging diagnostics.\cite{zweben-nf,maqueda,terry-php,zweben-php,terry-jnm2,terry-jnm1,zweben-php-acm,grulke-php,zweben-nf,agostini,cziegler-php,agostini-ppcf,kube-jnm,zweben-ppcf,fuchert-ppcf} During their radial propagation, the blob-like structures develop a steep front and a trailing wake, which can also be observed in numerical simulations of isolated blob structures\cite{bian-php,garcia-php-blob1,garcia-php-blob2,aydemir,kube-php-blob1,angus-php-blob1,kube-php-blob2,angus-php-blob2,madsen,gingell,kube-php-blob3,russell-php,wiesenberger,easy,halpern} and simulations of scrape-off layer turbulence.\cite{sarazin-jnm,garcia-prl-esel,bisai-php1,garcia-php-esel,garcia-ps-esel,fundamenski-nf,horacek-nf,russel-php,militello-ppcf,yan,bisai-php2} Conditional averaging of experimental measurement data have shown that large-amplitude fluctuations are well described by an exponential wave form.\cite{boedo-php1,rudakov-ppcf,boedo-php2,rudakov-nf,kirnev,antar-php3,garcia-ppcf-tcv1,garcia-jnm-tcv,garcia-nf-tcv,garcia-ppcf-tcv2,xu,horacek-asdex,militello,carralero-nf,boedo-php3,walkden,garcia-jnm,garcia-aps,garcia-nfl,theodorsen-ppcf,kube-ppcf,garcia-nme} This universal observation of front steepening in interchange motions of plasma filaments motivates the present study of exponential pulses.

The statistical properties of large-amplitude fluctuations in the scrape-off layer of tokamak plasmas have recently been elucidated by means of exceptionally long data time series under stationary plasma conditions.\cite{garcia-jnm,garcia-aps,garcia-nfl,theodorsen-ppcf,kube-ppcf,garcia-nme,theodorsen-nfl} From single-point measurements it has been demonstrated that there is an exponential distribution of both the peak amplitudes and the waiting times between them. Moreover, the average duration time does not depend on the amplitude and also appears to be robust against changes in plasma parameters.\cite{garcia-jnm-tcv,garcia-nf-tcv,garcia-ppcf-tcv2,garcia-jnm,garcia-aps,garcia-nfl,theodorsen-ppcf,kube-ppcf,garcia-nme,theodorsen-nfl} Based on these properties, a stochastic model for the plasma fluctuations has been developed by a super-position of uncorrelated exponential pulses.\cite{garcia-prl,garcia-php,theodorsen-php,kube-php,theodorsen-ps,garcia-phpl} The underlying assumptions and predictions of this model have recently been found to compare favorably with experimental measurements.\cite{garcia-jnm,garcia-aps,garcia-nfl,theodorsen-ppcf,kube-ppcf,garcia-nme,theodorsen-nfl}

In this contribution, the auto-correlation function and frequency spectrum are derived and discussed in detail for one- and two-sided exponential pulses. These are shown to be independent of the amplitude distribution of the pulses as well as the degree of pulse overlap and thereby the intermittency of the stochastic process. For constant pulse duration and a one-sided exponential pulse shape, the power spectral density has a Lorentzian shape, which is flat for low frequencies and a power law spectrum at high frequencies. For a two-sided exponential pulse shape, the power spectral density is the product of two Lorentzian functions. The power law tails at high frequencies are shown to result from discontinuities in the pulse function or its derivative. A distribution of pulse duration times is demonstrated to result in apparently longer correlation times but has no influence on the power law tail of the frequency spectrum.

This paper is organized as follows. In the following section the stochastic model describing fluctuations due to a super-position of uncorrelated pulses is presented. In \Secref{sec:moments} the mean and variance of the random variable are calculated and the intrinsic intermittency of the process is discussed. General expressions for the auto-correlation function and the power spectral density are derived in \Secref{sec:afcpsd}. The cases of one- and two-sided exponential pulse functions with constant pulse duration is considered in \Secref{sec:phiexp}. In \Secref{sec:powerlaw} the algebraic tail in the frequency spectra are demonstrated to result from the discontinuity in the pulse function or its derivative. A distribution of pulse durations is in \Secref{sec:ptau} shown to result in apparently longer correlation times but has no effect on the asymptotic power law tail in the frequency spectrum. The contribution of additional random noise is discussed in \Secref{sec:noise} and finally a discussion of the results and the conclusions are given in \Secref{sec:disc}. The probability density functions in the case of exponential and Laplace amplitude distributions are derived in Appendix \ref{sec:pdf}. A discussion of the relation between discontinuities in the pulse shape or its derivatvies and power law spectra is given in Appendix \ref{sec:gamma}. Finally, the role of additional random noise is discussed in Appendix \ref{sec:noise}.

\section{Stochastic model}\label{sec:model}

Consider a stochastic process given by the super-position of a random sequence of $K$ pulses in a time interval of duration $T$,\cite{garcia-prl,kube-php,theodorsen-php,garcia-php,theodorsen-ps,garcia-phpl,rice,parzen,pecseli,lowen}
\begin{equation}\label{shotnoise}
\Phi_K(t) = \sum_{k=1}^{K(T)} A_k\phi\left( \frac{t-t_k}{\tau_k} \right) ,
\end{equation}
where each pulse labeled $k$ is characterized by an amplitude $A_k$, arrival time $t_k$, and duration $\tau_k$, all assumed to be uncorrelated and each of them independent and identically distributed. The pulse arrival times $t_k$ are in the following assumed to be uniformly distributed on the time interval under consideration, that is, their probability density function is given by $1/T$. The pulse duration times $\tau_k$ are assumed to be randomly distributed with probability density $P_\tau(\tau)$, and the average pulse duration time is defined by
\begin{equation}\label{Ptaud}
\taud = \langle \tau \rangle = \int_0^\infty \rmd\tau\,\tau P_\tau(\tau) .
\end{equation}
Here and in the following, angular brackets denote the average of the argument over all random variables. The results presented here are independent of the distribution of pulse amplitudes $P_A(A)$, it is only assumed that the mean $\Aave$ and variance $\langle{A^2}\rangle$ are finite. The role of the pulse amplitude distribution is discussed further in Appendix \ref{sec:pdf}.

The pulse shape $\phi(\theta)$ is taken to be the same for all events in \Eqref{shotnoise}. This function is normalized such that
\begin{equation}\label{duration}
\int_{-\infty}^{\infty} \rmd\theta\,\abs{\phi(\theta)} = 1 .
\end{equation}
Thus, for constant duration each pulse contributes equally to the mean value of $\Phi_K(t)$. The integral of the $n$'th power of the pulse shape is defined as
\begin{equation}\label{pulseint}
I_n = \int_{-\infty}^{\infty} \rmd\theta\,\left[ \phi(\theta) \right]^n ,
\end{equation}
for positive integers $n$. It is assumed that $T$ is large compared with the range of values of $t$ for which $\phi(t/\tau)$ is appreciably different from zero, thus allowing to neglect end effects in a given realization of the process. Furthermore, the normalized auto-correlation function of the pulse function is defined by\cite{garcia-phpl}
\begin{equation}\label{rhophi}
\rho_\phi(\theta) = \frac{1}{I_2}\int_{-\infty}^{\infty} \rmd\chi\,\phi(\chi)\phi(\chi+\theta) ,
\end{equation}
and the Fourier transform of this gives the frequency spectrum,
\begin{equation}\label{varrhophi}
\Varrho_\phi(\vartheta) = \int_{-\infty}^\infty \rmd\theta\,\rho_\phi(\theta)\exp(-i\vartheta\theta) = \frac{1}{I_2}\,\abs{\varphi}^2(\vartheta) ,
\end{equation}
where the transform of the pulse function is defined by
\begin{equation}\label{varphi}
\varphi(\vartheta) = \int_{-\infty}^\infty \rmd\theta\,\phi(\theta)\exp(-i\vartheta\theta) .
\end{equation}
In this contribution, the auto-correlation function and the power spectral density for the process defined by \Eqref{shotnoise} will be investigated for the case of exponential pulses. The influence of various distributions of the pulse duration times as well as additive random noise will be explored. However, first the two lowest order moments of the process will be derived.

\section{Mean, variance and intermittency}\label{sec:moments}

The two lowest order moments of a realization of the stochastic process defined by \Eqref{shotnoise} can be calculated in a straight forward manner by averaging over all random variables.\cite{garcia-prl,kube-php,theodorsen-php,garcia-php,theodorsen-ps,garcia-phpl,rice,parzen,pecseli}

\subsection{Mean value}

Starting with the case of exactly $K$ events in a time interval with duration $T$, the mean value is given by
\begin{multline} \label{avephiK}
\langle{\Phi_K}\rangle = \int_{-\infty}^\infty \rmd A_1\,P_A(A_1) \int_0^\infty \rmd\tau_1 P_\tau(\tau_1) \int_0^T \frac{\rmd t_1}{T}
\\
\cdots \int_{-\infty}^\infty \rmd A_K\,P_A(A_K) \int_0^\infty \rmd\tau_K\,P_\tau(\tau_K) \int_0^T \frac{\rmd t_K}{T} \sum_{k=1}^K A_k\phi\left( \frac{t-t_k}{\tau_k} \right) ,
\end{multline}
using that the pulse arrival times are uniformly distributed. Neglecting end effects by taking the integration limits for the arrival times $t_k$ in \Eqref{avephiK} to infinity, the mean value of the signal follows directly,
\begin{equation}
\langle{\Phi_K}\rangle = \taud I_1 \ave{A} \frac{K}{T} .
\end{equation}
Here a change of integration variable defined by $\theta=(t-t_k)/\tau_k$ has been made, giving
\begin{equation}
\int_0^\infty \rmd\tau_k\,P_\tau(\tau_k) \int_{-\infty}^\infty \rmd t_k \phi\left( \frac{t-t_k}{\tau_k} \right) = \int_0^\infty \rmd\tau_k\,\tau_k P_\tau(\tau_k) \int_{-\infty}^\infty \rmd\theta\,\phi(\theta) = \taud I_1 .
\end{equation}
Taking into account that the number of pulses $K$ is also a random variable and averaging over this as well gives the mean value for the stationary process,
\begin{equation} \label{phiave}
\ave{\Phi} = \frac{\taud}{\tauw}\,I_1\ave{A} ,
\end{equation}
where $\tauw=T/\langle{K}\rangle$ is the average pulse waiting time. For a non-negative pulse function, $I_1=1$, the mean value of the process is given by the average pulse amplitude and the ratio of the average pulse duration and waiting times. The mean value vanishes both for anti-symmetric pulse shapes, $I_1=0$, and for pulse amplitude distributions with vanishing mean, $\ave{A}=0$. It should also be noted that the mean values does not depend on the distribution of pulse duration times.

\subsection{Variance}

The variance can similarly be calculated by averaging the square of the random varible. From the definition of $\Phi_K(t)$ it follows that
\begin{equation}
\Phi_K^2(t) = \sum_{k=1}^K \sum_{\ell=1}^K A_k A_\ell \phi\left( \frac{t-t_k}{\tau_k} \right)\phi\left( \frac{t-t_\ell}{\tau_\ell} \right) .
\end{equation}
Averaging this over pulse amplitudes as well as duration and arrival times for fixed $t$ and $K$ gives
\begin{multline}
\langle{\Phi_K^2}\rangle = \int_{-\infty}^\infty \rmd A_1\,P_A(A_1) \int_0^\infty \rmd\tau_1\,P_\tau(\tau_1) \int_0^T \frac{\rmd t_1}{T}
\\
\cdots \int_{-\infty}^\infty \rmd A_K\,P_A(A_K) \int_0^\infty \rmd\tau_K\,P_\tau(\tau_K) \int_0^T \frac{\rmd t_K}{T} \sum_{k=1}^K \sum_{\ell=1}^K A_kA_\ell\phi\left( \frac{t-t_k}{\tau_k} \right)\phi\left( \frac{t-t_\ell}{\tau_\ell} \right) .
\end{multline}
There are two contributions to the variance from the double sum. When $k\neq\ell$ there are $K(K-1)$ terms with the value
\begin{multline}
\int_{-\infty}^\infty \rmd A_k\,P_A(A_k) \int_0^\infty \rmd\tau_k\,P_\tau(\tau_k) \int_0^T \frac{\rmd t_k}{T} \,A_k\phi\left( \frac{t-t_k}{\tau_k} \right)
\\
\times \int_{-\infty}^\infty \rmd A_\ell\,P_A(A_\ell) \int_0^\infty \rmd\tau_\ell\,P_\tau(\tau_\ell) \int_0^T \frac{\rmd t_\ell}{T} A_\ell\phi\left( \frac{t-t_\ell}{\tau_\ell} \right) ,
\end{multline}
while for $k=\ell$ there are $K$ terms given by the integral
\begin{equation}
\int_{-\infty}^\infty \rmd A_k\,P_A(A_k) \int_0^\infty \rmd\tau_k\,P_\tau(\tau_k) \int_0^T \frac{\rmd t_k}{T} A_k^2 \phi^2\left( \frac{t-t_k}{\tau_k} \right) .
\end{equation}
Neglecting end effects due to the finite duration of individual pulses by extending the integration limits for $t_k$ and $t_\ell$ to infinity, the variance for large $T$ in the case of exactly $K$ pulses is given by
\begin{equation}
\langle{\Phi_K^2}\rangle = \taud I_2 \langle{A^2}\rangle\,\frac{K}{T} + \taud^2 I_1^2 \ave{A}^2\,\frac{K(K-1)}{T^2} .
\end{equation}
By averaging over all realizations where the number of pulses $K$ is statistically distributed gives
\begin{equation}
\langle{\Phi^2}\rangle = \frac{\taud}{\tauw}\,I_2\langle{A^2}\rangle + \Phiave^2 ,
\end{equation}
where $\langle{K(K-1)}\rangle=\langle{K}\rangle^2$ has been assumed. This is an exact relation for a Poisson distribution of the number of pulses, and approximate when there is a large number of pulses in each realization of the process. It follows that the standard deviation or root mean square (rms) value of the random variable is given by
\begin{equation} \label{phivariance}
\Phirms^2 = \frac{\taud}{\tauw}\,I_2\langle{A^2}\rangle .
\end{equation}
Thus, the absolute fluctuation level is large when there is significant overlap of pulse events, that is, for long pulse durations and short pulse waiting times. As for the mean value given above, the variance does not depend on the distribution of the pulse duration times.

\subsection{Intermittency}

The ratio of the average pulse duration and waiting times,
\begin{equation}
\gamma = \frac{\taud}{\tauw} ,
\end{equation}
determines the degree of pulse overlap in the stochastic process. When $\gamma$ is small, pulses generally appear isolated in realizations of the process, resulting in a strongly intermittent signal where most of the time is spent at small values. When $\gamma$ is large, there is significant overlap of pulses and realizations of the process resemble random noise. Indeed, it can be demonstrated that the probability density function for the random variable $\Phi_K(t)$ approaches a normal distribution in the limit of infinitely large $\gamma$, independently of the pulse shape and amplitude distribution.\cite{garcia-prl,garcia-php}

\begin{figure}
\includegraphics[width=10cm]{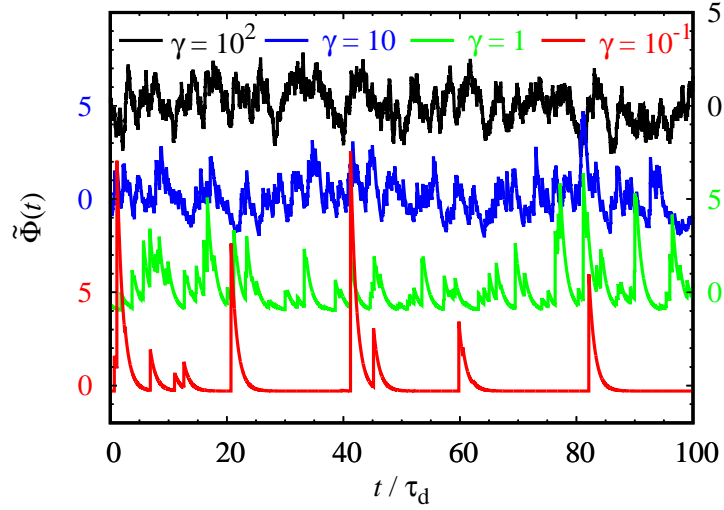}
\caption{Realizations of the stochastic process for a one-sided exponential pulse shape with constant duration $\taud$ and exponentially distributed pulse amplitudes. The degree of pulse overlap is determined by the intermittency parameter $\gamma=\taud/\tauw$.}
\label{fig:Phiraw}
\end{figure}

For a non-zero mean value of the process, the relative fluctuation level is given by
\begin{equation}
\frac{\Phirms^2}{\Phiave^2} = \frac{1}{\gamma}\frac{I_2}{I_1^2}\frac{\langle{A}^2\rangle}{\Aave^2} .
\end{equation}
This is large for long pulse waiting times and short pulse durations. In Appendix \ref{sec:pdf} it is shown that also the skewness and flatness moments become large for small values of $\gamma$. In the following, the rescaled variable with zero mean and unit standard deviation will be frequently considered,
\begin{equation}\label{Phiwt}
\Phiwt(t) = \frac{\Phi-\Phiave}{\Phirms} .
\end{equation}
Some realizations of this process are presented in \Figref{fig:Phiraw} for one-sided exponential pulses with constant duration and exponentially distributed amplitudes. It is clearly seen that the signal is strongly intermittent for small values of $\gamma$, while pulse overlap for large values of $\gamma$ makes the signals resemble random noise. Further description of the intermittency effects in this process is given in Appendix \ref{sec:pdf} and are discussed in \Refs{garcia-prl,garcia-php,kube-php,theodorsen-php,theodorsen-ps,garcia-phpl}.

\section{Correlations and spectra}\label{sec:afcpsd}

In the same way as the variance was calculated above, the auto-correlation function for $\Phi_K(t)$ can be calculated by a straight forward average over all random variables and the power spectral density is then given by a transformation to the frequency domain.

\subsection{Auto-correlation function}

Considering first the signal $\Phi_K(t)$ defined by \Eqref{shotnoise}, the auto-correlation function for a time lag $r$ is given by a double sum,
\begin{multline}\label{PhiKcorr}
\langle{\Phi_K(t)\Phi_K(t+r)}\rangle = \int_{-\infty}^\infty \rmd A_1\,P_A(A_1) \int_0^\infty \rmd\tau_1\,P_\tau(\tau_1) \int_0^T \frac{\rmd t_1}{T} \cdots
\\
\int_{-\infty}^\infty \rmd A_K\,P_A(A_K) \int_0^\infty \rmd\tau_K\,P_\tau(\tau_K) \int_0^T \frac{\rmd t_K}{T} \sum_{k=1}^K\sum_{\ell=1}^K A_k\phi\left( \frac{t-t_k}{\tau_k} \right) A_\ell\phi\left( \frac{t-t_\ell+r}{\tau_\ell} \right) .
\end{multline}
There are again two contributions to the double sum, comprising $K(K-1)$ terms when $k\neq\ell$ given by
\begin{equation}
\aveA^2\sum_{\substack{k,\ell=1\\ k\neq\ell}}^K \int_0^\infty \rmd\tau_k\,P_\tau(\tau_k) \int_0^T \frac{\rmd t_k}{T}\,\phi\left( \frac{t-t_k}{\tau_k} \right) \int_0^\infty \rmd\tau_\ell\,P_\tau(\tau_\ell) \int_0^T \frac{\rmd t_\ell}{T}\,\phi\left( \frac{t-t_\ell+r}{\tau_\ell} \right) ,
\end{equation}
and $K$ terms when $k=\ell$,
\begin{equation}
\langle{A^2}\rangle \sum_{k=1}^K \int_0^\infty \rmd\tau_k\,P_\tau(\tau_k) \int_0^T\frac{\rmd t_k}{T}\phi\left( \frac{t-t_k}{\tau_k} \right)\phi\left( \frac{t-t_k+r}{\tau_k} \right) .
\end{equation}
Neglecting end effects due to the finite pulse duration by taking the integration limits for $t_k$ and $t_\ell$ to infinity for the case of exactly $K$ pulse events results in
\begin{equation}
\langle{\Phi_K(t)\Phi_K(t+r)}\rangle =
\taud^2 I_1^2 \aveA^2\,\frac{K(K-1)}{T^2} 
+ I_2 \langle{A^2}\rangle\,\frac{K}{T}\,\int_0^\infty \rmd\tau\,\tau P_\tau(\tau) \rho_\phi(r/\tau) .
\end{equation}
Averaging over the number of pulses occurring in the interval of duration $T$, it follows that the auto-correlation function for the stationary process is given by\cite{garcia-phpl}
\begin{equation} \label{Rphi}
R_{\Phi}(r) = \Phiave^2 + \Phirms^2\,\frac{1}{\taud}\int_0^\infty \rmd\tau\,\tau P_\tau(\tau) \rho_\phi(r/\tau) ,
\end{equation}
where $\rho_\phi(\theta)$ is the normalized auto-correlation function for the pulse function defined by \Eqref{rhophi}. The expression of the auto-correlation function is simplified by considering the rescaled variable defined by \Eqref{Phiwt},
\begin{equation} \label{RPhiwt}
R_{\Phiwt}(r) = \frac{1}{\taud}\int_0^\infty \rmd\tau\,\tau P_\tau(\tau) \rho_\phi(r/\tau) .
\end{equation}
It is emphasized that this expression for the auto-correlation function does not rely on a Poisson distribution of the number of pulse events. The auto-correlation function is determined by the pulse shape through $\rho_\phi(\theta)$ and the probability density function for the pulse duration times.

Equation~\eqref{Rphi} emphasizes the role of the pulse shape in determining the auto-correlation function for the process. However, the time lag $r$ in the integrand can be transfered to the distribution function for the pulse durations times by the change of variables $\theta=\abs{r}/\tau$, which gives the alternative formulation
\begin{equation}
R_{\Phiwt}(r) = \frac{1}{\taud}\int_0^\infty \frac{\rmd\theta}{\theta^2}\frac{r}{\theta}\,\abs{r} P_\tau(\abs{r}/\theta)\rho_\phi(\theta) , 
\end{equation}
where the probability density function for the pulse duration times is normalized such that
\begin{equation}
\int_0^\infty \rmd\tau\,P_\tau(\tau) = \int_0^\infty \frac{\rmd\theta}{\theta^2}\,\abs{r} P_\tau(\abs{r}/\theta) = 1 .
\end{equation}
In the case of constant pulse duration, the latter is given by the degenerate distribution $P_\tau(\tau)=\delta(\tau-\taud)$, where $\delta$ is the delta distribution. The auto-correlation function is then $R_{\Phiwt}(r)=\rho_\phi(r/\taud)$, that is, it is simply given by the normalized auto-correlation function of the individual pulses in the proecss.

\subsection{Power spectral density}

From the auto-correlation function given above, the power spectral density follows directly by a Fourier transformation to the frequency domain,\cite{garcia-phpl}
\begin{equation}
\Omega_\Phi(\omega) = \int_{-\infty}^\infty \rmd r\,R_\Phi(r)\exp(-i\omega r) = 2\pi\Phiave^2\delta(\omega) + \Phirms^2\frac{1}{\taud}\int_0^\infty \rmd\tau\,\tau^2 P_\tau(\tau)\Varrho_\phi(\tau\omega) ,
\end{equation}
where $\omega$ is the angular frequency and $\Varrho_\phi(\vartheta)$ is the Fourier transform of the normalized auto-correlation function defined by \Eqref{varrhophi}. Here the first term on the right hand side of the second equality is a zero-frequency contribution from the mean value of the random variable. The expression for the power spectral density is simplified by considering the rescaled random variable defined by \Eqref{Phiwt},
\begin{equation}\label{psdphiwt}
\Omega_{\wt{\Phi}}(\omega) = \frac{1}{\taud}\int_0^\infty \rmd\tau\,\tau^2 P_\tau(\tau) \Varrho_\phi(\tau\omega) .
\end{equation}
It should be noted that this frequency spectrum is independent of the amplitude distribution $P_A$ and does not depend on the intermittency parameter $\gamma$. The latter property is evidently due to the assumption of uncorrelated pulses. Moreover, the above expression is not restricted to a Poisson distribution for the number of pulses $K(T)$. The only assumptions made are that the pulse arrival times have a uniform distribution and that the two lowest order moments for the process are finite. In the special case of constant pulse duration, the expression for the power spectral density become $\Omega_{\Phiwt}(\omega)=\taud\Varrho_\phi(\taud\omega)$, that is, the spectrum is simply determined by that of the pulse function $\phi(\theta)$.

By the simple change of variables, $\vartheta=\tau\abs{\omega}$, \Eqref{psdphiwt} for the power spectral density can be written in the alternative form
\begin{equation}\label{psd-power}
\Omega_{\Phiwt}(\omega) = \frac{1}{\taud}\int_0^\infty \frac{\rmd\vartheta}{\abs{\omega}}\frac{\vartheta^2}{\omega^2}\,P_\tau(\vartheta/\abs{\omega})\Varrho_\phi(\vartheta) , 
\end{equation}
where the probability density function for the pulse duration times is normalized such that
\begin{equation}
\int_0^\infty \rmd\tau\,P_\tau(\tau) = \int_0^\infty \frac{\rmd\vartheta}{\abs{\omega}}\,P_\tau(\vartheta/\abs{\omega}) = 1 .
\end{equation}
Equation \eqref{psd-power} is suggestive of a power law spectrum for sufficiently broad distributions of pulse durations. In general, the power spectral density is determined by both the pusle shape and the distribution function for pulse durations.\cite{garcia-phpl}

\section{Exponential pulses}\label{sec:phiexp}

The above expressions for the auto-correlation function and the power spectral density were derived for an arbitrary pulse function under the assumption that the two lowest order moments exist. In this section, the properties of the stochastic process will be investigated for the case of one- and two-sided exponential pulses with constant duration.

\subsection{One-sided exponential pulse}

Consider first the standard case of an one-sided exponential pulse shape given by\cite{garcia-prl,garcia-php,kube-php} % FIX ORDER
\begin{equation}\label{phi-exp}
\phi(\theta) = \Theta(\theta)\exp(-\theta) ,
\end{equation}
where $\Theta(\theta)$ is the unit step function defined by
\begin{equation}\label{unitstep}
\Theta(\theta) =
\begin{cases}
1 , & \theta \geq 0 ,
\\
0 , & \theta < 0 .
\end{cases}
\end{equation}
The integral of the $n$'th power of the pulse shape is in this case given by $I_n=1/n$. The auto-correlation function for the pulse shape and its transform then become
\begin{subequations}\label{rhovar-exp}
\begin{align}%\label{rho-exp}
\rho_\phi(\theta) & = \exp(-\abs{\theta}) ,\label{rhophi-exp}
\\
\Varrho_\phi(\vartheta) & = \frac{2}{1+\vartheta^2} .\label{varrhophi-exp}
\end{align}
\end{subequations}
In the case of a super-position of uncorrelated, one-sided exponential pulses with constant duration, it follows that the auto-correlation function is a symmetric exponential function,\cite{garcia-php,kube-php}
\begin{equation}\label{acf-exp}
R_{\Phiwt}(r) = \exp\left( - \frac{\abs{r}}{\taud} \right) ,
\end{equation}
while the power spectral density is a Lorentzian function,
\begin{equation}\label{psd-exp}
\frac{1}{2\taud}\,\Omega_{\Phiwt}(\omega) = \frac{1}{1+\taud^2\omega^2} .
\end{equation}
This well-known frequency spectrum is flat for low frequencies and has a power law tail for high frequencies. The auto-correlation function and the power spectral density for one-sided exponential pulses are presented in \Figsref{fig:acf-exp-lambda} and \ref{fig:psd-exp-lambda}, respectively.

\subsection{Two-sided exponential pulse}

\begin{figure}
\includegraphics[width=10cm]{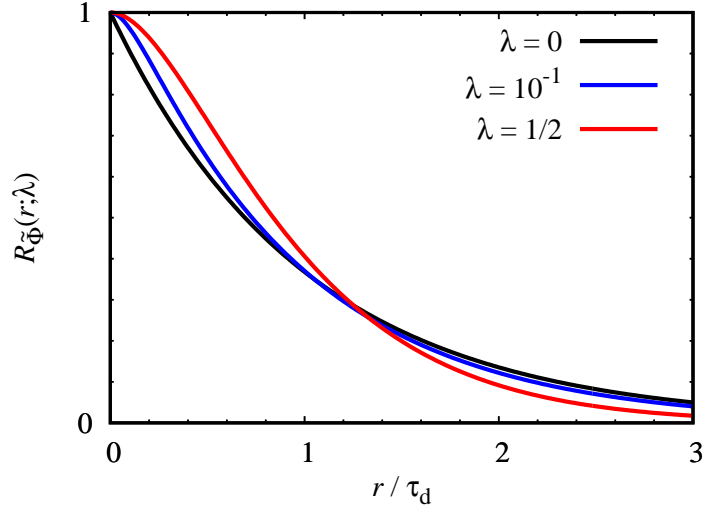}
\caption{Auto-correlation function for a super-position of uncorrelated, two-sided exponential pulses with constant duration for various values for the asymmetry parameter $\lambda$. The case $\lambda=0$ corresponds to a one-sided exponential pulse shape.}
\label{fig:acf-exp-lambda}
\end{figure}

The case of a one-sided exponential pulse function is readily generalized to a pulse shape that is continuous by introducing a finite pulse rise time,\cite{theodorsen-ppcf,theodorsen-php,garcia-aps}
\begin{equation}\label{phi-dexp}
\phi(\theta;\lambda) = \Theta(-\theta)\exp\left( \frac{\theta}{\lambda} \right) + \Theta(\theta)\exp\left( -\frac{\theta}{1-\lambda} \right) ,
\end{equation}
where the pulse asymmetry parameter $\lambda$ is restricted to the range $0<\lambda<1$. For $\lambda<1/2$ the pulse rise time, $\lambda\taud$, is faster that than the decay time, $(1-\lambda)\taud$, and the pulse shape is symmetric in the case $\lambda=1/2$. The integral of the $n$'th power of the pulse function is the same as for the one-sided exponential pulse discussed above, $I_n=1/n$. Thus, while the probability density function and the moments are the same, the auto-correlation function for the pulse function and its transform are altered by the finite rise time,
\begin{subequations}\label{rhovar-dexp}
\begin{align}\label{rho-dexp}
\rho_\phi(\theta;\lambda) & = \frac{1}{1-2\lambda}\left[ (1-\lambda)\exp\left( -\frac{\abs{\theta}}{1-\lambda} \right) - \lambda\exp\left( -\frac{\abs{\theta}}{\lambda} \right) \right] ,
\\
\Varrho_\phi(\vartheta;\lambda) & = \frac{2}{[1+(1-\lambda)^2\vartheta^2][1+\lambda^2\vartheta^2]} . \label{varrho-dexp}
\end{align}
\end{subequations}
In the case of constant pulse duration, the auto-correlation function for a super-position of uncorrelated pulses is therefore given by\cite{theodorsen-ppcf}
\begin{equation}\label{acf-dexp}
R_{\Phiwt}(r;\lambda) = \frac{1}{1-2\lambda}\left[ (1-\lambda)\exp\left( -\frac{\abs{r}}{(1-\lambda)\taud} \right) - \lambda\exp\left( -\frac{\abs{r}}{\lambda\taud} \right) \right] .
\end{equation}
For a one-sided exponential pulse shape, the auto-correlation function given by \Eqref{acf-exp} has an exponential decay from zero time lag. With the finite rise time for a two-sided exponential pulse shape, the correlation function is flat for zero time lag. This is particularly clear in the symmetric case $\lambda=1/2$, for which the auto-correlation function given by
\begin{equation}
R_{\Phiwt}(r;1/2) = \left( 1+ \frac{2\abs{r}}{\taud} \right)\exp\left( - \frac{2\abs{r}}{\taud} \right) ,
\end{equation}
has a parabolic shape for small time lags. The auto-correlation function is presented in \Figref{fig:acf-exp-lambda} for various values of $\lambda$.

\begin{figure}
\includegraphics[width=10cm]{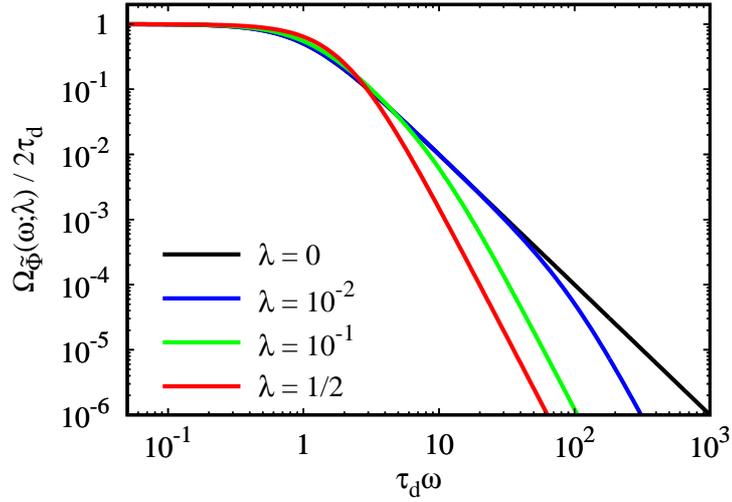}
\caption{Power spectral density for a super-position of uncorrelated, two-sided exponential pulses with constant duration for various values for the asymmetry parameter $\lambda$. The case $\lambda=0$ corresponds to a one-sided exponential pulse shape.}
\label{fig:psd-exp-lambda}
\end{figure}

The power spectral density for a super-position of two-sided exponential pulses with constant duration is given by
\begin{equation}\label{psd-dexp}
\frac{1}{2\taud}\,\Omega_{\Phiwt}(\omega;\lambda) = \frac{1}{\left[ 1+(1-\lambda)^2\taud^2\omega^2 \right]\left[ 1+\lambda^2\taud^2\omega^2 \right]} .
\end{equation}
This spectrum is flat for low frequencies and has a steep power law scaling for high frequencies, which is particularly clear for a symmetric pulse shape,
\begin{equation}
\frac{1}{2\taud}\,\Omega_{\Phiwt}(\omega;1/2) = \frac{1}{\left( 1+\frac{1}{4}\taud^2\omega^2 \right)^2} .
\end{equation}
However, for $\lambda\ll1$ or $1-\lambda\ll1$, the frequency spectrum resembles a broken power law with an intermediate power law scaling regime given by $1/(\taud\omega)^2$. When the pulse rise and fall times are comparable, the spectrum will appear curved over a large range of frequencies when presented in a double-logarithmic plot. This is clearly seen in \Figref{fig:psd-exp-lambda}, where the frequency spectrum is presented for various values of the asymmetry parameter $\lambda$.

\section{Power law spectra}\label{sec:powerlaw}

In this section a general relation is derived for the power spectral density of the process defined by \Eqref{shotnoise} and its derivative in the case of constant pulse duration. From this it is demonstrated that the power law tails in the frequency spectra for exponential pulses are due to a discontinuity in the pulse function or its derivative. Further discussions on the relation between discontinuities and algebraic tails in the power spectral densities are given in Appendix \ref{sec:gamma}.

\subsection{Derivatives and discontinuities}

Consider the normalized time derivative of the stochastic process given by \Eqref{shotnoise} in the case of constant pulse duration $\taud$,
\begin{equation}
\Psi_K(t) = c_\phi \taud\,\frac{\rmd\Phi_K}{\rmd t} = \sum_{k=1}^{K(T)} A_k \psi\left(\frac{t-t_k}{\taud}\right) ,
\end{equation}
where the pulse function for the process $\Psi_K(t)$ is related to $\phi(\theta)$ by
\begin{equation}
\psi(\theta) = c_\phi\,\frac{\rmd\phi}{\rmd\theta} .
\end{equation}
Here a normalization constant $c_\phi$ has been introduced in order for the pulse shape $\psi(\theta)$ to satisfy the requirement given by \Eqref{duration} for the stochastic process,
\begin{equation}
\int_{-\infty}^{\infty} \rmd\theta\abs{\psi(\theta)} = 1 .
\end{equation}
The constant $c_\phi$ clearly depends on the pulse function $\phi(\theta)$. The integral of the $n$'th power of the derivative of this pulse function is defined as
\begin{equation}
J_n = \int_{-\infty}^{\infty} \rmd\theta[\psi(\theta)]^n ,
\end{equation}
Provided that the mean value and variance exist for both the process $\Phi_K(t)$ and its derivative, it is straight forward to show that $J_2\Varrho_\psi(\vartheta)=I_2 c_\phi^2 \vartheta^2\Varrho_\phi(\theta)$. Thus, the power spectral densities for the stochastic process and its derivative in the case of constant pulse duration are related by
\begin{equation}\label{psd-ddt}
J_2 \Omega_{\Psiwt}(\omega) = I_2 c_\phi^2\taud^2\omega^2 \Omega_{\Phiwt}(\omega) .
\end{equation}
As will be shown in the following, this relation provides an explanation of the power law tails in the frequency spectra for the one- and two-sided exponential pulse shapes described by \Eqsref{psd-exp} and \eqref{psd-dexp}, respectively. In order to demonstrate this, the frequency spectra for two elementary pulse functions will be reviewed.

\subsection{Box pulse}\label{sec:box}

Consider a pulse shape given by the unit box function on the interval $\abs{\theta}\leq1/2$, defined by
\begin{equation}\label{phi-box}
\Pi(\theta) = \Theta\left(\theta+\frac{1}{2} \right) - \Theta\left( \theta-\frac{1}{2} \right) ,
\end{equation}
where $\Theta(\theta)$ is the unit step function given by \Eqref{unitstep}. The normalized auto-correlation function $\varrho_\phi(\theta)$ for this pulse shape equals the unit triangle function defined by
\begin{equation}\label{phi-triangle}
\Lambda(\theta) =
\begin{cases}
1-\abs{\theta}, & \abs{\theta} \leq 1 ,
\\
0, & \abs{\theta} > 1 .
\end{cases}
\end{equation}
The Fourier transform of this function gives $\Varrho_\phi(\vartheta)=2(1-\cos{\vartheta})/\vartheta^2$. A super-position of uncorrelated box pulses gives a process with mean value $\Phiave=\gamma\Aave$ and variance $\Phirms^2=\gamma\langle{A^2}\rangle$, since $I_n=1$ for the unit box function. In the case of constant pulse duration, the auto-correlation function for this process is given by the unit triangle function and the power spectral density is
\begin{equation}\label{psd-box}
\frac{1}{2\taud}\,\Omega_{\Phiwt}(\omega) = \frac{1-\cos(\taud\omega)}{\taud^2\omega^2} .
\end{equation}
This frequency spectrum has the same algebraic tail for high frequencies as the process with a super-position of one-sided exponential pulse functions given by \Eqref{psd-exp}.

The derivative of the box pulse in \Eqref{phi-box} can be calculated by means of the theory of functionals and results in delta distributions associated with the two discontinuities at $\theta=\pm1/2$ for the unit box function,
\begin{equation}\label{psi-box}
\psi(\theta) = \frac{1}{2}\frac{\rmd\Pi}{\rmd\theta} = \frac{1}{2}\,\delta\left(\theta+\frac{1}{2}\right) - \frac{1}{2}\,\delta\left(\theta-\frac{1}{2}\right) .
\end{equation}
It should be noted that the integral $J_2$ is not defined for this pulse function. However, the unnormalized auto-correlation function can be calculated and gives a sum of three delta distributions,
\begin{equation}\label{rhopsi-box}
\frac{1}{4}\int_{-\infty}^\infty \rmd\chi\,\psi(\chi)\psi(\chi+\theta) = - \delta(\theta+1) + 2\delta(\theta) - \delta(\theta-1) .
\end{equation}
The Fourier transform gives a flat frequency spectrum with periodic oscillations,
\begin{equation}\label{varrho-box}
\int_{-\infty}^{\infty} \rmd\theta\,\exp(-i\vartheta\theta) \left[ - \delta(\theta+1) + 2\delta(\theta) - \delta(\theta-1) \right] = 2(1-\cos{\vartheta}) .
\end{equation}
The flat spectrum obviously arises from the transform of the delta pulses in the derivative of the unit box function. Based on the relation between the frequency spectrum for a stationary process and its derivative discussed above, the $1/(\taud\omega)^2$ power law tail for the spectrum given by \Eqref{psd-exp} is attributed to the discontinuity at $\theta=0$ for the one-sided exponential pulse function defined by \Eqref{phi-exp}.

\subsection{Triangle pulse}\label{sec:triangle}

In order to elucidate the origin of the steep power spectral density at high frequencies for a two-sided exponential pulse function, consider the unit triangle pulse on the interval $\abs{\theta}\leq1$ defined by \Eqref{phi-triangle}. In this case the integral over the $n$'th power of the pulse function is $I_n=2/(1+n)$ and the normalized auto-correlation function for the unit triangle pulse is
\begin{equation}
\rho_\phi(\theta) = 
\begin{cases}
\frac{1}{4}\left[ 4-3(2-\abs{\theta})\theta^2 \right] , & 0 \leq \abs{\theta} < 1 ,
\\
\frac{1}{4}(2-\abs{\theta})^3 , & 1 \leq \abs{\theta} < 2 ,
\\
0 , & 2 \leq \abs{\theta} .
\end{cases}
\end{equation}
For a super-position of uncorrelated triangle pulses the mean value and variance are given by $\Phiave=\gamma\Aave$ and $\Phirms^2=2\gamma\langle{A^2}\rangle/3$, respectively, and the frequency spectrum in the case of constant pulse duration is
\begin{equation}\label{psd-triangle}
\frac{1}{2\taud}\,\Omega_{\Phiwt}(\omega) = \frac{12\sin^4(\taud\omega/2)}{\taud^4\omega^4} .
\end{equation}
This has the same algebraic tail at high frequencies as the power spectral density for a super-position of two-sided exponential pulses given by \Eqref{psd-dexp}.

In order to reveal the origin of this steep power law spectrum, consider the derivative of the triangle pulse function which has discontinuities at $\theta=0$ and $\pm1$,
\begin{equation}
\psi(\theta) = \frac{1}{2}\frac{\rmd\Lambda}{\rmd\theta} = \frac{1}{2}\left[ \Theta(\theta+1) - 2\Theta(\theta) + \Theta(\theta-1) \right] .
\end{equation}
The variance for this pulse function is $J_2=1/2$. It is straight forward to calculate the normalized auto-correlation function and its Fourier transform, which gives $\Varrho_\psi(\vartheta)=8\sin^4(\vartheta/2)/\vartheta^2$ and therefore a power spectral density for a super-position of such pulses given by
\begin{equation}
\frac{1}{2\taud}\,\Omega_{\Psiwt}(\omega) = \frac{4\sin^4(\taud\omega/2)}{\taud^2\omega^2} ,
\end{equation}
confirming the general relation given by \Eqref{psd-ddt}. Based on the relation between the frequency spectrum for a stationary process and its derivative discussed above, the steep $1/(\taud\omega)^4$ algebraic tail for the power spectral density in \Eqref{psd-dexp} is attributed to the break point at $\theta=0$ for the two-sided exponential pulse function given by \Eqref{phi-dexp}. Similarly, the apparent broken power law with an intermediate $1/(\taud\omega)^2$ scaling regime for either $\lambda\ll1$ or $1-\lambda\ll1$ results from the abrupt change in the pulse shape in these cases. However, when $0<\lambda<1$ the frequency spectrum eventually breaks into the steep $1/(\taud\omega)^4$ power law tail for sufficiently high frequencies, as clearly seen in \Figref{fig:psd-exp-lambda}.

\section{Distribution of pulse durations}\label{sec:ptau}

In this section, some particular cases of pulse duration distributions which result in closed analytical expressions for the auto-correlation function and power spectral density in the case of a one-sided exponential pulse shape are discussed in detail.

\subsection{Rayleigh distribution}

In the case of a Rayleigh distribution of pulse durations, $\taud P_\tau(\tau)=(\pi\tau/2\taud)\exp(-\pi\tau^2/2\taud^2)$, the power spectral density is given by
\begin{equation}
\frac{1}{2\taud}\,\Omega_{\Phiwt}(\omega) = \frac{1}{\taud^2\omega^2}
\left[ 1 - \frac{\pi}{4\taud^2\omega^2}\,\exp\left( \frac{\pi}{4\taud^2\omega^2} \right)\Gamma\left( 0,\frac{\pi}{4\taud^2\omega^2} \right) \right] ,
\end{equation}
where $\Gamma$ here denotes the incomplete Gamma function. This spectrum has the asymptotic limits
\begin{subequations}
\begin{gather}
\lim_{\taud\abs{\omega}\rightarrow0} \frac{\pi}{4}\frac{\Omega_{\Phiwt}(\omega)}{2\taud} = 1 ,
\\
\lim_{\taud\abs{\omega}\rightarrow\infty} \taud^2\omega^2\,\frac{\Omega_{\Phiwt}(\omega)}{2\taud} = 1  .
\end{gather}
\end{subequations}
The latter is notably the same power law tail as for the case of constant pulse duration.

\subsection{Gamma distribution}

A general probability density function for the pulse duration times is given by the Gamma distribution,
\begin{equation}
\taud P_\tau(\tau;s) = \frac{s^s}{\Gamma(s)}\left( \frac{\tau}{\taud} \right)^{s-1}\exp\left( - \frac{s\tau}{\taud} \right) ,
\end{equation}
with scale parameter $\taud$ and shape parameter $s$. For small values of $s$, there is a high probability for small duration times. For $s=1$, $P_\tau$ is an exponential distribution, while for $s=2$ it is similar to a Rayleigh distribution except that is has an exponential tail for long duration times, $\taud P_\tau(\tau;2)=(4\tau/\taud)\exp(-2\tau/\taud)$. For large values of $s$, $P_\tau$ resembles a normal distribution and the limit $s\rightarrow\infty$ corresponds to the case with constant pulse duration.

\begin{figure}
\includegraphics[width=10cm]{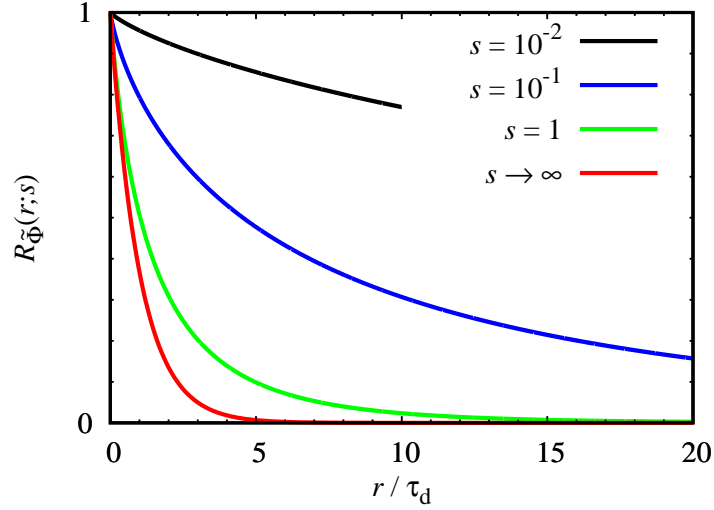}
\caption{Auto-correlation function for a super-position of uncorrelated, one-sided exponential pulses with a Gamma distribution of pulse durations with mean value $\taud$ and shape parameters $s$. The limit $s\rightarrow\infty$ corresponds to the case with constant pulse duration.}
\label{fig:acf-gamma}
\end{figure}

For the one-sided exponential pulse shape defined by \Eqref{phi-exp}, the auto-correlation function is given by
\begin{equation}
R_{\Phiwt}(r;s) = \frac{2}{s\Gamma(s)}\,\abs{sr/\taud}^{(s+1)/2} \mathcal{K}_{s+1}(2\abs{sr/\taud}^{1/2}) ,
\end{equation}
where $\mathcal{K}$ is the modified Bessel function of the second kind. The tail of this function is a stretched exponential,
\begin{equation}
\lim_{r/\taud\rightarrow\infty} \frac{s\Gamma(s)}{\pi^{1/2}} \frac{\exp\left( 2\abs{sr/\taud}^{1/2} \right)}{\abs{sr/\taud}^{(2s+1)/4}}\,R_{\Phiwt}(r;s) = 1 .
\end{equation}
The auto-correlation function is presented in \Figref{fig:acf-gamma} for various values of the shape parameter $s$. For small $s$, there is a pronounced tail in the correlation function. The power spectral density is given by hypergeometric functions and has the following asymptotic limits,
\begin{gather}
\lim_{\taud\abs{\omega}\rightarrow0} \frac{\Omega_{\Phiwt}(\omega;s)}{2\taud} = \frac{1+s}{s} ,
\\
\lim_{\taud\abs{\omega}\rightarrow\infty} \taud^2\omega^2\,\frac{\Omega_{\Phiwt}(\omega;s)}{2\taud} = 1 .
\end{gather}
This clearly shows that the energy in the low-frequency part of the spectrum is large for small values of the shape parameter, while the power law tail for high frequencies is the same as for the case with constant pulse duration.

\subsection{Uniform distribution}

\begin{figure}
\includegraphics[width=10cm]{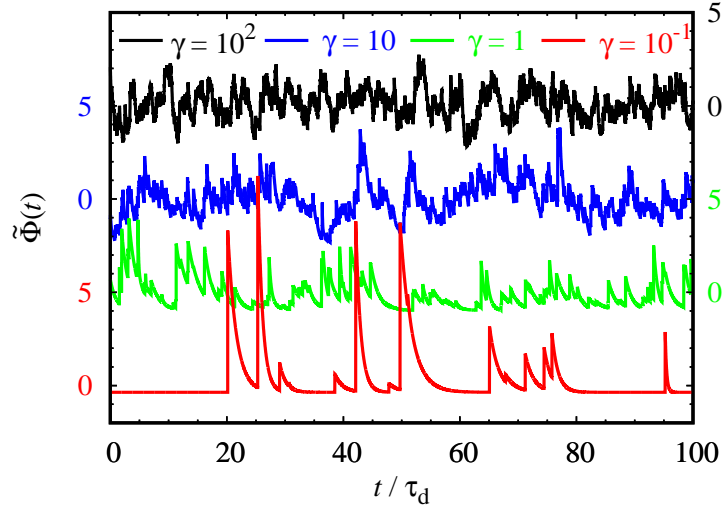}
\caption{Realizations of the stochastic process for a one-sided exponential pulse shape with a uniform distribution of pulse durations and exponentially distributed pulse amplitudes. The degree of pulse overlap is determined by the intermittency parameter $\gamma=\taud/\tauw$.}
\label{fig:Phiraw-unif}
\end{figure}

Consider finally the case with a uniform distribution of pulse durations, $\taud P_\tau(\tau;s)=1/2s$, for duration times in the range $1-s<\tau/\taud<1+s$ and the parameter $s$ in the range $0<s\leq1$. The limit $s\rightarrow0$ corresponds to the case with constant pulse duration. Some realizations of this process are presented in \Figref{fig:Phiraw-unif} for one-sided exponential pulses, exponentially distributed amplitudes and $s=1$. For this distribution, the power spectral density is given by
\begin{equation}
\frac{1}{2\taud}\,\Omega_{\Phiwt}(\omega;s) = \frac{2s\taud\omega - \text{atan}[(1+s)\taud\omega]+\text{atan}[(1-s)\taud\omega]}{2s\taud^3\omega^3} ,
\end{equation}
which has the asymptotic limits
\begin{gather}
\lim_{\taud\abs{\omega}\rightarrow0} \frac{\Omega_{\Phiwt}(\omega;s)}{2\taud} = \frac{3+s^2}{3} ,
\\
\lim_{\taud\abs{\omega}\rightarrow\infty} \taud^2\omega^2\,\frac{\Omega_{\Phiwt}(\omega;s)}{2\taud} = 1 ,
\end{gather}
again showing that the power law spectrum for high frequencies is just the same as for the case with constant pulse duration.

\begin{figure}
\includegraphics[width=10cm]{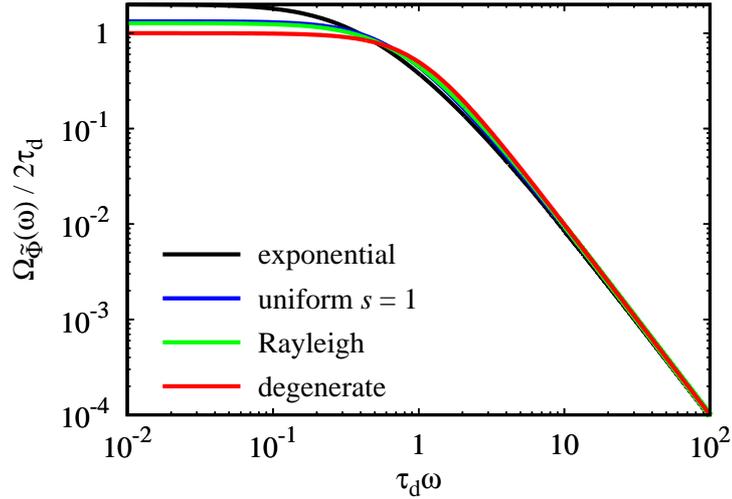}
\caption{Power spectral density for a super-position of uncorrelated, one-sided exponential pulses for various duration time distributions. The degenerate case corresponds to constant pulse duration.}
\label{fig:psd-compare}
\end{figure}

\begin{figure}
\includegraphics[width=10cm]{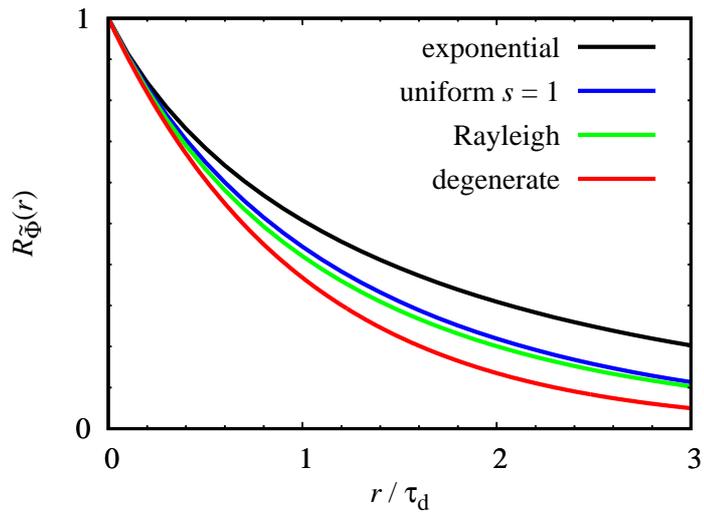}
\caption{Auto-correlation function for a super-position of uncorrelated, one-sided exponential pulses for various duration time distributions. The degenerate case corresponds to constant pulse duration.}
\label{fig:acf-compare}
\end{figure}

The power spectral densities for a super-position of uncorrelated, one-sided exponential pulses in the cases of exponential, uniform, Rayleigh and degenerate distributions of pulse duration times are compared in \Figref{fig:psd-compare}. The asymptotic power law tail for high frequencies is the same for all these distributions, while there are slight variations in the spectral power at low frequencies. The corresponding auto-correlation functions are presented in \Figref{fig:acf-compare}, showing that a broad distribution of pulse durations leads to apparently longer correlation times. An analysis of the effect of a distribution of pulse durations in the case of two-sided exponential pulses gives qualitatively similar results as those presented above, expect for the steeper $1/(\taud\omega)^4$ algebraic tail for high frequencies.

\section{Discussion and conclusions}\label{sec:disc}

A reference model for intermittent fluctuations in physical systems has here been extended to include a random distribution of pulse duration times. This is demonstrated to modify the auto-correlation function and power spectral density, which in general are determined by the pulse shape and the distribution of duration times. However, the intermittency parameter which determines the degree of pulse overlap does not influence the auto-correlation function nor the power spectral density. Conversely, it has here been shown that the distribution of pulse durations does not influence the probability density function for the process, and thus neither the moments. These general results hold for arbitrary pulse amplitude distributions and pulse shapes.

In this work, particular attention has been given to exponential pulses since they provide a favorable description of large-amplitude bursts measured in the boundary region of magnetically confined plasmas. In the case of a one-sided exponential pulse shape, the auto-correlation function is an exponential and the power spectral density is a Lorentzian function which is flat for low frequencies and has a $1/(\taud\omega)^2$ power law tail for high frequencies. By considering the time derivative of the process, this power law tail is here demonstrated to result from the discontinuity in the one-sided exponential pulse shape.

For a two-sided exponential pulse shape, the power spectral density is the product of two Lorentzian functions and has a steep $1/(\taud\omega)^4$ algebraic tail for high frequencies. This is shown to result from the break point at the peak of the pulse function. For a strongly asymmetric pulse shape, the frequency spectrum resembles a broken power law with an intermediate power law scaling regime given by $1/(\taud\omega)^2$. This is obviously due to the abrupt change in the pulse shape. However, since this is a continuous function, the power spectral density eventually breaks into the steep $1/(\taud\omega)^4$ algebraic tail for sufficiently high frequencies.

A random distribution of pulse durations is shown to result in long apparent correlation times but has no influence on the asymptotic power law tail in the frequency spectrum. The former is possibly related to the many reports on long-range dependence of plasma fluctuations in the boundary of magnetically confined plasmas.\cite{pedrosa-prl,carreras-php,rhodes,antar-php1,antar-php2} Under the assumption of self-similarity, the estimate of scaling coefficients sensitively depends on the presence of low-frequency trends and oscillations which typically are present in measurement data. It is to be noted that the stochastic model presented here obviously does not possess any self-similarity properties.

Additional random noise can preclude the asymptotic scaling regimes, and if sufficiently large, the frequency spectrum may appear curved over a large range of frequencies. This underlies the crucial importance of accurate measurements and long data time series in order to reliably estimate such scaling relationships. The predicted auto-correlation function and frequency spectrum for two-sided exponential pulses presented here has recently been shown be in excellent agreement with measurement data from the outboard mid-plane region of several tokamak experiments.\cite{theodorsen-nfl,garcia-nme,theodorsen-ppcf}

Finally, it is to be noted that many previous investigations of pink or $1/\omega$ noise in physical systems has referred to the stochastic process with a one-sided exponential pulse shape investigated here. In the literature this is well-known as \emph{shot noise} or a \emph{filtered Poisson process}.\cite{rice,parzen,pecseli,lowen} The effect of a distribution of pulse durations has commonly been included by performing an ensamble average of the frequency spectrum derived for constant pulse duration, here given by \Eqref{psd-exp}. For a broad uniform distribution of pulse duration times, this leads to the erroneous but widely cited conclusion that this process is capable of producing $1/\omega$-type spectra.\cite{vanziel1,pre,butz,hooge,vanziel2,kogan} On the other hand, it has been established that truncated power law pulse shapes may lead to pink noise, with such processes commonly referred to as fractal shot noise.\cite{lowen}

\appendix

\section{Probability densities}\label{sec:pdf}

The characteristic function for a sum of independent random variables is the product of their individual characteristic functions. For a super-position of exactly $K$ uncorrelated pulses in a time interval of duration $T$, the conditional probability density function for the process defined by \Eqref{shotnoise} is thus given by
\begin{equation}%\label{prbK}
P_\Phi(\Phi | K) = \frac{1}{2\pi}\int_{-\infty}^{\infty} \rmd u\,\exp(-iu\Phi) \langle{\exp(iu\phi_k)}\rangle^K ,
\end{equation}
where the angular brackets as usual denote an average over all random variables of $\phi_k=A_k\phi((t-t_k)/\tau_k$,
\begin{equation}
\langle{\exp(iu\phi_k)}\rangle = \int_{-\infty}^{\infty} \rmd A_k P_A(A_k) \int_{0}^{\infty} \rmd\tau_k P_\tau(\tau_k) \int_0^T \frac{\rmd t_k}{T}\, \exp\left( iu A_k\phi\left( \frac{t-t_k}{\tau_k} \right) \right) .
\end{equation}
Assuming that the number of pulses $K(T)$ is Poisson distributed,
\begin{equation} \label{poisson}
P_K(K|T) = \frac{1}{K!}\left(\frac{T}{\tauw}\right)^K\exp\left(-\frac{T}{\tauw} \right) ,
\end{equation}
the probability density function for the random variable $\Phi$ is given by
\begin{equation}
P_\Phi(\Phi) = \sum_{K=0}^{\infty} P_\Phi(\Phi|K) P_K(K|T) = \frac{1}{2\pi}\int_{-\infty}^{\infty} \rmd u\,\exp( - iu\Phi )C_\Phi(u) .
\end{equation}
where the characteristic function is defined by
\begin{equation}
{C_\Phi(u)} = \exp\left( \frac{T}{\tauw}\,\langle\exp(iu\phi_k)\rangle - \frac{T}{\tauw} \right) . 
\end{equation}
Taking the limit of integration for $t_k$ to infinity and making a change of integration variable defined by $\theta=(t-t_k)/\tau_k$ results in
\begin{equation}
\ln{C_\Phi(u)} = \gamma \int_{-\infty}^{\infty} \rmd\theta \int_{-\infty}^{\infty} \rmd A P_A(A) \left[ \exp\left( iuA\phi(\theta) \right) - 1 \right] = \gamma \int_{-\infty}^{\infty} \rmd\theta \left[ C_A(u\phi(\theta)) - 1 \right]
\end{equation}
where $C_A(u)$ is the characteristic function for the amplitude distribution $P_A(A)$. Thus, the probability density function is determined by amplitude distribution and the pulse shape and is not affected by the distribution of pulse durations. This is in agreement with the two lowest order moments derived in \Secref{sec:moments}.

For some particular cases of pulse shapes and amplitude distributions, the probability density function can be calculated in closed form. Consider first the case of an exponential distribution of pulse amplitudes,
\begin{equation}
P_A(A) = \frac{1}{\Aave}\,\exp\left( - \frac{A}{\Aave} \right) ,
\end{equation}
where $\Aave$ is the mean pulse amplitude and $P_A$ is defined only for positive amplitudes $A$. In this case the raw amplitude moments are given by $\langle{A^n}\rangle=n!\Aave^n$ and the mean value for the process is $\Phiave=\gamma\Aave$ and the variance is $\Phirms^2=\gamma\Aave^2/2$. For both one- and two-sided exponential pulses, the characteristic function is $C_\Phi(u)=(1+iu\Aave)^\gamma$ and the probability density function for the random variable is a Gamma distribution\cite{garcia-prl,garcia-php,theodorsen-php}
\begin{equation}
\Phiave P_\Phi(\Phi) = \frac{\gamma}{\Gamma(\gamma)}\left( \frac{\gamma\Phi}{\Phiave} \right)^{\gamma-1}\exp\left( - \frac{\gamma\Phi}{\Phiave} \right) .
\end{equation}
Thus, the skewness and flatness moments are $S_\Phi=2/\gamma^{1/2}$ and $F_\Phi=3+6/\gamma$, revealing the strong intermittency of the process in the case of weak pulse overlap. For large $\gamma$ the skewness and excess flatness moments both vanish, consistent with a normal distribution of the fluctuations which arise in this limit.\cite{garcia-prl,garcia-php}

Allowing both positive and negative pulse amplitudes, the Laplace amplitude distribution with vanishing mean is of particular interest,
\begin{equation}
P_A(A) = \frac{1}{2^{1/2}\Arms}\,\exp\left( - \frac{2^{1/2}\abs{A}}{\Arms} \right) ,
\end{equation}
where $\Arms$ is the standard deviation, $\langle{A^2}\rangle=\Arms^2$. 
%The odd moments for this distribution vanish, while for the even moments $\langle{A^{2n}}\rangle=(2n)!(\Arms/2^{1/2})^{2n}$ for positive integers $n$. 
For this symmetric distribution, both the mean value and the skewness moment for the random variable vanish, $\Phiave=0$ and $S_\Phi=0$. For both one- and two-sided exponential pulses the variance of the random variable is $\Phirms^2=\gamma\Arms^2/2$ and the probability density function is given by\cite{theodorsen-ppcf}
\begin{equation}
\Phirms P_\Phi(\Phi) = \frac{\gamma^{1/2}}{\pi^{1/2}\Gamma(\gamma/2)}\left( \frac{\gamma^{1/2}\abs{\Phi}}{2\Phirms} \right)^{(\gamma-1)/2}\mathcal{K}_{(\gamma-1)/2}\left( \frac{\gamma^{1/2}\abs{\Phi}}{\Phirms} \right) ,
\end{equation}
where $\mathcal{K}$ denotes the modified Bessel function of the second kind.
It follows that the flatness moment is $F_\Phi=3+6/\gamma$, which is the same as for exponentially distributed amplitudes discussed above. %
% SUMMARIZE: LARGE INTERMITTENCY, NORMAL LIMIT FOR LARGE GAMMA
% For large values of the intermittency parameter $\gamma$,

\section{Gamma pulses}\label{sec:gamma}

In general, the Fourier transform of the $m$'th derivative of a pulse function $\phi(\theta)$ is given by $(i\vartheta)^m\varphi$, where $\varphi(\vartheta)$ is the Fourier amplitude as defined by \Eqref{varphi}. Consider a pulse function that is smooth for all $\theta$ except at $\theta=0$, where the $m$'th derivative is dominated by a delta distribution due to a discontinuity in the $m-1$'th derivative, $\rmd^m\phi/\rmd\theta^m\sim\delta(\theta)$. The Fourier transform of the $m$'th derivative therefore gives a flat spectrum, $(i\vartheta)^m\varphi(\vartheta)\sim1$. Thus, the power spectral density for the pulse function is anticipated to have a power law asymptote given by $\abs{\varphi}^2\sim1/\vartheta^{2m}$. This is a heuristic description of the power law scaling in the frequency spectra for exponential pulses described in \Secref{sec:phiexp}.

As an illustrative example for which exact results can be calculated, consider here a Gamma shaped pulse function defined by
\begin{equation}
\phi(\theta;s) = \Theta(\theta)\,\frac{s^s\theta^{s-1}}{\Gamma(s)}\,\exp(-s\theta) ,
\end{equation}
for positive integer values of the shape parameter $s$. For $s=1$, this is a one-sided exponential pulse which is discontinuous at $\theta=0$. In the limit $s\rightarrow\infty$, the pulse shape approaches a smooth Gaussian function. More generally, the pulse function has a discontinuity in the $s-1$ derivative and the Fourier transform of the $s$ derivative will be dominated by a delta distribution. Indeed, the case $s=1$ gives the familiar result
\begin{equation}
\frac{\rmd\phi}{\rmd\theta}(\theta,1) = \delta(\theta) - \Theta(\theta)\exp(-\theta) ,
\end{equation}
and the power spectral density scales as $\Varrho_\phi(\vartheta)\sim1/\vartheta^2$.
Similarly, for $s=2$ the pulse function has a break point at $\theta=0$ and the second derivative of the pulse function is dominated by a delta distribution,
\begin{equation}
\frac{\rmd^2\phi}{\rmd\theta^2}(\theta,2) = 4\delta(\theta) + \Theta(\theta)(\theta-1)\exp(-2\theta) ,
\end{equation}
resulting in a power spectral density that which scales as $1/\vartheta^4$ for high frequencies, similar to the case of the unit triangle pulse discussed in \Secref{sec:triangle}.

The power spectral density for a super-position of Gamma pulses with constant duration can be calculated explicitly and is given by
\begin{equation}
\frac{1}{2\taud}\,\Omega_{\Phiwt}(\omega) = \frac{\Gamma(s)}{\Gamma(s-1/2)} \frac{\pi^{1/2}s^{2s-1}}{(s^2+\taud^2\omega^2)^s}
\end{equation}
This frequency spectrum has the asymptotic limit
\begin{equation}
\lim_{\taud\abs{\omega}\rightarrow\infty} \taud^{2s}\omega^{2s}\,\frac{\Omega_{\Phiwt}(\omega)}{2\taud} = \frac{\Gamma(s)\pi^{1/2}s^{2s-1}}{\Gamma(s-1/2)} ,
\end{equation}
as expected from the discontinuity in the $s-1$'th derivative of the pulse function and the general relation given by \Eqref{psd-ddt}. This example clearly demonstrates the intimate connection between a discontinuity in the pulse function or its derivatives and algebraic tail in the power spectral density for high frequencies.

\section{Additive noise}\label{sec:noise}

In many relevant cases there will be additional noise to the stochastic process defined by \Eqref{shotnoise}. Consider here the case where statistically independent and random noise is added to the process comprised by a super-position of uncorrelated pulses,\cite{garcia-nme,theodorsen-ps}
\begin{equation}
\Psi_K(t) = \Phi_K(t) + \sigma N(t) ,
\end{equation}
where $N$ is a normally distributed process with vanishing mean and unit standard deviation. It is straight forward to shown that the mean value and variance for the process $\Psi_K(t)$ are $\Psiave=\gamma I_1\Aave$ and $\Psirms^2=(1+\epsilon)\gamma\langle{A^2}\rangle$, respectively. The signal to noise ratio is the inverse of the ratio of the variance for the two independent processes,
\begin{equation}
\epsilon = \frac{\sigma^2}{\Phirms^2} .
\end{equation}
%
%stationary probability density function for the process $\Psi_K(t)$ is a convolution of a Gamma and a normal distribution with mean value $\gamma\Aave$ and variance $\Psirms^2=(1+\epsilon)\gamma I_2\langle{A^2}\rangle$, where the ratio of the variance for the two independent processes are defined by
%\begin{equation}
%\epsilon = \frac{\sigma^2}{\Phirms^2} .
%\end{equation}
Introducing the rescaled signals as defined by \Eqref{Phiwt}, it follows that the auto-correlation function for the summed processes is given by
\begin{equation}
R_{\Psiwt}(r) = \frac{ R_{\Phiwt}(r) + \epsilon R_{\Nwt}(r) }{1+\epsilon} ,
\end{equation}
where $R_{\Nwt}(r)$ the auto-correlation function for the rescaled noise process.

\begin{figure}
\includegraphics[width=10cm]{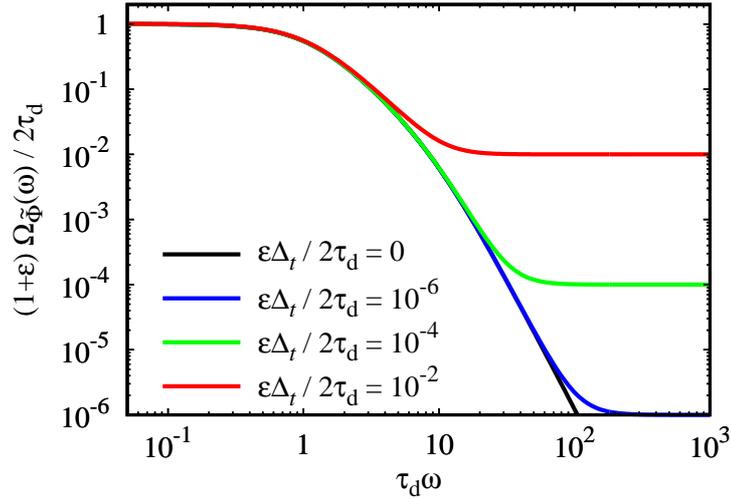}
\caption{Power spectral density for a super-position of uncorrelated, two-sided exponential pulses with $\lambda=1/10$ and constant duration for various values for the signal to noise ratio.}
\label{fig:psd-exp-noise}
\end{figure}

Delta-correlated noise can be written as the integrated increments of a Wiener process,
\begin{equation}
N(t) = \frac{1}{\triangle_t^{1/2}} \int_t^{t+\triangle_t} \rmd W(r) ,
\end{equation}
where $\triangle_t$ is the assumed constant sampling time. It should be noted that for such a discretely sampled signal, no power will be recorded for frequencies higher than the Nyquist frequency which is given by $\omega_\text{max}=\pi/\triangle_t$. The auto-correlation function for this process is
\begin{equation}
R_N(r) = \left( 1 - \frac{\abs{r}}{\triangle_t} \right)\Theta(\triangle_t-\abs{r}) ,
\end{equation}
and the power spectral density for the noise process is given by
\begin{equation}
\frac{1}{2\triangle_t}\,\Omega_N(\omega) = \int_{-\infty}^\infty \rmd r\,\exp(-i\omega r)R_N(r) = \frac{1-\cos(\triangle_t\omega)}{\triangle_t^2\omega^2} .
\end{equation}
A power series expansion of the cosine function gives the power spectral density for the rescaled process $\Psiwt_K(t)$,
\begin{equation}
\Omega_{\Psiwt}(\omega) = \frac{1}{1+\epsilon}\left[ \Omega_{\Phiwt}(\omega) + \epsilon\triangle_t \right] ,
\end{equation}
where the second term inside the square parenthesis is the flat spectrum resulting from the delta-correlated noise. For the process with two-sided exponential pulses with constant duration, the power spectral density is thus given by
\begin{equation}
\frac{1}{2\taud}\,\Omega_{\Psiwt}(\omega) = \frac{1}{1+\epsilon}\left\{ \frac{1}{\left[ 1+(1-\lambda)^2\taud^2\omega^2 \right]\left[ 1+\lambda^2\taud^2\omega^2 \right]} + \frac{\epsilon\triangle_t}{2\taud} \right\} .
\end{equation}
This frequency spectrum is presented in \Figref{fig:psd-exp-noise} for various values of the signal to noise ratio. It is clearly seen that for significant noise levels, the power spectrum appears curved over a large range of frequencies and the underlying power law asymptote may be precluded by the noise.

\section*{Acknowledgements}

This work was supported with financial subvention from the Research Council of Norway under grant 240510/F20. The authors acknowledge the generous hospitality of the MIT Plasma Science and Fusion Center where this work was conducted.

\end{document}